\documentclass[11.pt]{article}

\usepackage{latexsym}
\usepackage{epsfig}
\usepackage{color}
\usepackage{caption}
\def\R{ {\rm R \kern -.31cm I \kern .15cm}}
\def\C{ {\rm C \kern -.15cm \vrule width.5pt \kern .12cm}}
\def\Z{ {\rm Z \kern -.27cm \angle \kern .02cm}}
\def\N{ {\rm N \kern -.26cm \vrule width.4pt \kern .10cm}}
\def\1{{\rm 1\mskip-4.5mu l} }
\def\lsim{\raise0.3ex\hbox{$<$\kern-0.75em\raise-1.1ex\hbox{$\sim$}}}
\def\gsim{\raise0.3ex\hbox{$>$\kern-0.75em\raise-1.1ex\hbox{$\sim$}}}

\def\beq{\begin{equation}}   \def\eeq{\end{equation}}
\def\bea{\begin{eqnarray}}  \def\eea{\end{eqnarray}}

\DeclareGraphicsExtensions{.eps}

\def\lsim{\raise0.3ex\hbox{$<$\kern-0.75em\raise-1.1ex\hbox{$\sim$}}}
\def\gsim{\raise0.3ex\hbox{$>$\kern-0.75em\raise-1.1ex\hbox{$\sim$}}}

\begin{document}

\title{\bf High multiplicity pp events and $J/\psi$ production at LHC}

\vskip 8. truemm
\author{\bf E. G. Ferreiro and C. Pajares}
\vskip 5. truemm

\date{}
\maketitle

\begin{center}
\small{
  Departamento de F{\'\i}sica de Part{\'\i}culas and IGFAE, Universidad de
  Santiago de Compostela, \\
  15782 Santiago de Compostela, Spain}
\end{center}
\vskip 5. truemm

\begin{abstract}
We discuss the dependence of $J/\psi$ production on the charged particle multiplicity in proton-proton collisions at LHC energies. 
We show that, in the framework of parton saturation or string interaction models, the hard $J/\psi$ production exhibits a significant growth with the multiplicity, which is stronger than linear in the high density domain.
This departure from linearity, that   
should affect any hard observable, 
applies for high multiplicity proton-proton collisions in the central rapidity region and 
is a consequence of the 
parton saturation or the strong interaction
among colour ropes
that 
take place at LHC energies. 
Our assumption, the existence of coherence effects present in proton-proton collisions at high energy, can also be checked
by studying the particular shape of the probability distribution associated to the $J/\psi$ production.

\end{abstract}
\vskip 3 truecm

\hspace*{\parindent}
\pagestyle{myheadings}
RHIC \cite{Back:2004je,Adcox:2004mh,Adams:2005dq,Arsene:2004fa} and LHC \cite{Schukraft:2011cz,Steinberg:2011dj,Wyslouch:2011zz} data on heavy-ion collisions have shown several 
important features which indicate the formation of a high density partonic medium with characteristic properties as the low shear viscosity and high opacity. Since the energy density achieved in high multiplicity events produced in $pp$ collisions at LHC is comparable to the reached density in CuCu central collisions or AuAu peripheral collisions at $\sqrt{s_{NN}}=200$ GeV, it is 
pertinent 
to wonder about the possibility to obtain a similar high density medium which would be reflected in experimental 
observables,
similar to heavy-ion collisions.  An illustration of this would be the predicted ridge structure \cite{Cunqueiro:2008uu,Brogueira:2009nj} observed by CMS collaboration \cite{Khachatryan:2010gv}
 in $pp$ collisions. 
Also, the eventuality that other observables, as long range rapidity correlations \cite{deDeus:2010id,Strikman:2011ar,Werner:2010ss}, energy loss \cite{Vogel}, or the elliptic flow \cite{Bautista:2009my,CasalderreySolana:2009uk}, are measurable in $pp$ collisions, 
has been reckoned with in different frameworks. 

%
%
%
%
%
%

We address here to the $J/\psi$ production in high multiplicity $pp$ collisions.
We will show that the rise of $J/\psi$ production in the highest multiplicity events observed by 
the ALICE collaboration \cite{Abelev:2012rz}
 can be naturally explained as a consequence of string interaction or parton saturation.
This feature is particularly important, since, due to the absence of nuclear effects in this case, it can be use as a baseline in order to disentangle different mechanism that are expected in heavy-ion collisions, as the $J/\psi$ suppression due to 
sequential dissociation \cite{sequential}
 or the $J/\psi$ enhancement due to the recombination of uncorrelated $c$ and $\bar{c}$ quarks 
\cite{recombination}.

\vskip 0.35cm

Our main assumption, shared by many of the mentioned approaches, is the fact that, 
in high-energy hadronic collisions, all projectiles, be it
protons or nuclei, have finite spatial extension and thus
collide generically at finite impact parameter by means of elementary parton-parton collisions.
We may consider the color ropes or flux tubes - {\it strings} - as the fundamental
variables of our description. They are formed in each parton-parton collision and they constitute
the elementary sources of particle production.
In this string framework, the number of parton-parton collisions is reflected as the number of produced strings, $N_s$.
These strings have non-negligible transverse size, of the order of $0.2 \div 0.3$ fm and different space-time rapidities, 
and they can interact - {\it overlap}~-, so reducing the effective number of sources,
in particular in which concerns soft particle production.

Consider now the hard particle production. The number of initially produced $J/\psi$, $n_{J/\psi}$, 
can be taken as proportional to the number of collisions, in analogy to any hard process. 
In the string-like models, this number corresponds to the number of produced strings, $N_s$. 
On the other hand, the rapidity multiplicity distribution $dN/d\eta$ --mainly soft-- 
is not proportional to the number of collisions, but mostly to the number of participants. This reduction can be considered as a consequence of shadowing \cite{Ferreiro:2008wc}, parton saturation \cite{Kharzeev:2008nw} or string interactions -~{\it percolation}~-~\cite{Armesto:1996kt}. 
%

In the string percolation approach, the multiplicity distribution is given by
\beq
\frac{dN}{d\eta} = F(\rho) N_s \mu_1
\label{eq1}
\eeq
where $\mu_1$ corresponds to the multiplicity of a single string in the rapidity range of interest, $N_s$ is the number of produced strings and $F(\rho)$ corresponds to the 
damping 
factor induced by the string interaction, 
\beq
F(\rho)=\sqrt{\frac{1-e^{-\rho}}{\rho}} \, .
\label{eq2}
\eeq
Note that, within the damping factor, $1-e^{-\rho}$ represents the fraction of the total area that is covered by strings.
The interaction among strings and the consequent damping factor is a function of the 
the string density,
$\rho=\frac{N_s \sigma_0}{\sigma}$, where  $\sigma_0$ is the transverse area of one string,
$\sigma_0=\pi r_0^2$, $r_0\sim 0.25$ fm, and $\sigma$ corresponds to the transverse area of the collision. 
So the existence of many strings --directly related to the number of available partons-- effectively screens 
the charged particle multiplicities, which agrees qualitatively with the concept of saturation.

On the other hand, assuming the proportionality between the number of produced $J/\psi$ and the number of 
collisions,
\beq
\frac{n_{J/\psi}}{<n_{J/\psi}>}=\frac{N_s}{<N_s>} \, ,
\label{eq3}
\eeq
it is possible to obtain the relation between the charged particle multiplicity and the number of produced $J/\psi$, 
that, accordingly to eqs. (\ref{eq2}) and (\ref{eq3}), will obey
\beq
\frac{\frac{dN}{d\eta}}{<\frac{dN}{d\eta}>} = \left ( \frac{n_{J/\psi}}{<n_{J/\psi}>} \right )^{1/2} 
\left [ \frac{1-e^{-\frac{n_{J/\psi}}{<n_{J/\psi}>}<\rho>}}{1-e^{-<\rho>}} \right ]^{1/2} \, {\rm where}\, 
<\rho>=<N_s> \frac{\sigma_0}{\sigma}\, .
\label{eq4}
\eeq

At low multiplicities, where the number of strings $<N_s>$ is small, the above equation gives rise to the linear 
dependence 
\beq
\frac{n_{J/\psi}}{<n_{J/\psi}>}=\frac{\frac{dN}{d\eta}}{<\frac{dN}{d\eta}>} \, .
\label{eq5}
\eeq
On the contrary, at high multiplicities, the bracket in the right hand side of eq. (\ref{eq4}) can be approximated by 
$<\rho>^{-1/2}$. One obtains in this case 
\beq
\frac{n_{J/\psi}}{<n_{J/\psi}>}=<\rho> \left ( \frac{\frac{dN}{d\eta}}{<\frac{dN}{d\eta}>} \right )^2 \, .
\label{eq6}
\eeq
Thus the linear dependence obtained previously for low multiplicities, eq. (\ref{eq5}), changes to an squared dependence when high multiplicity events are at play.

\vskip 0.35cm
In order to compare with the available $pp$ experimental data, we take $\sigma=\sigma_{inel}^{pp}=70$ mb 
\cite{Antchev:2011zz} as the transverse area of the collision. The number of strings, $<N_s>$, can be obtained from the SFM code \cite{Amelin:2001sk}, a Monte Carlo code based on the quark gluon string model, similar to
the Dual Parton Model \cite{Capella:1992yb} or EPOS \cite{Pierog:2009zt}.
Moreover, the value of $N_s$ can also be calculated analytically \cite{DiasdeDeus:2005sq},
$<N_s> = b+ (2-b) \left(\frac{s}{s_t}\right)^{\lambda}$,
where $\lambda= 0.2 \div 0.3$, $b=1.37$ and the low energy threshold $\sqrt{s_t}= 10$ GeV.
We obtained, for the central rapidity region, $<N_s>=16$, while in the forward rapidity region the number of strings is smaller, $<N_s>=8$. The reason for this difference is the fact that, while the long strings --stretched between valence quarks and diquarks of the colliding protons-- cover most of the rapidity range thus contributing to both central and forward rapidity production, the short strings --stretched 
between sea quarks and antiquarks-- are mostly created in the central rapidity region only.

In Fig. 1 we show our results for both the central and forward rapidity range, together with the experimental data from the ALICE Collaboration \cite{Abelev:2012rz}.
\begin{figure*}[htb!]
\begin{center}
\vskip -3cm
\includegraphics[width=0.75\textwidth]{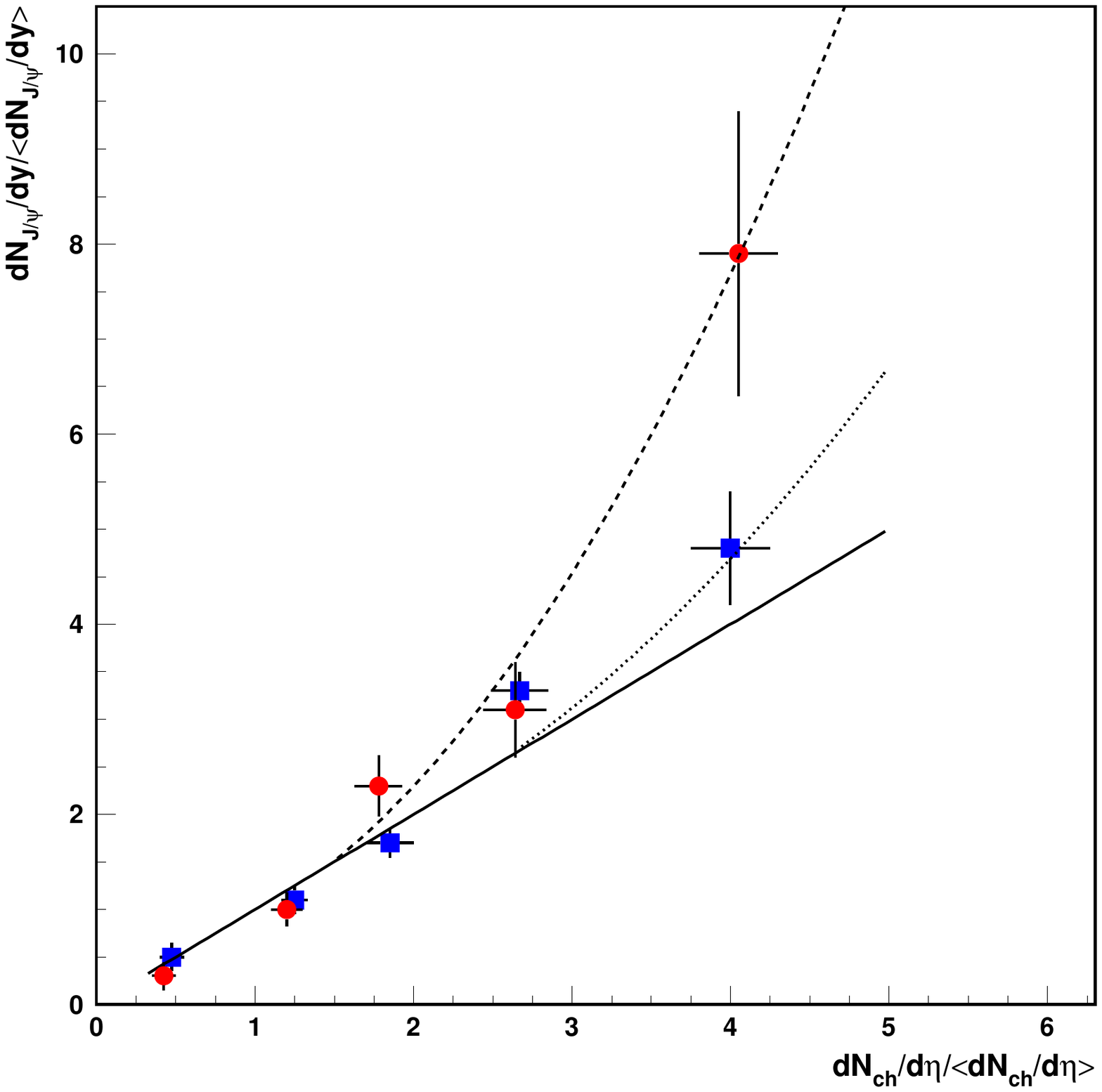}
\end{center}
\vskip -3cm
\caption{
Our results for the central ($|y|<0.9$, dashed line) and forward ($2.5 < y < 4$, dotted line) rapidity range, together with the experimental data 
for the central (circles) and forward (squares) rapidity regions} from the ALICE Collaboration \cite{Abelev:2012rz}. The linear behaviour
(solid line) is also plotted.
\label{fig1}
\end{figure*}
We observe a good agreement. The main uncertainty of our evaluation concerns the value of $<N_s>$, which can induce an uncertainty in the $<\rho>$ value of 20\%, of the same order of the last experimental 
point error.
Moreover, it would be most interesting to measure $\frac{n_{J/\psi}}{<n_{J/\psi}>}$ at larger multiplicity, 
$\frac{\frac{dN}{d\eta}}{<\frac{dN}{d\eta}>} \simeq 8 \div 10$ to check the behaviour of data.
Note that for this high $pp$ multiplicities, the values of the parton densities $\rho$ are comparable to those obtained in 
CuCu central collisions or AuAu peripheral collisions at RHIC energies, where $J/\psi$ melting is observed. If experimental data were significantly lower than our curve, this would be an indication of $J/\psi$ suppression in absence of any nuclear effect.

\vskip 0.35cm
Our assumption, the existence of coherence effects present in $pp$ collisions at high energy, can also be checked by studying the probability distribution associated to the $J/\psi$ production, $P_c(n)$. This distribution is given by
\beq
P_c(n)= \frac{n_{J/\psi}}{<n_{J/\psi}>} P(n)\, ,
\label{eq7}
\eeq
where $P(n)$ corresponds to the minimum bias probability distribution. 
The above equation refers to the well-established universal relation 
between minimum bias distribution $P$ and the multiplicity distribution $P_c$, associated to the production of a rare
event $c$ \cite{salga,DiasdeDeus:1998du}.
This formula is valid in any model of particle production based on 
the superposition of the contributions from elementary partonic interactions
--as it is the case in most of the multiple scattering models--, 
when the kind of events trigerred 
is shadowed only by events of the same kind, and not for the whole of events. 
In addition to this requirement, these events must have a small cross section.
In other words, one can say that the rare events are those 
produced only by one of the elementary interactions,
the probability of
it occurring twice in one
collision being negligible.
This is the case if one is triggering on a heavy particle like the $J/\psi$, 
and it would also be the case when studying $\Upsilon$ production or
multiplicity distributions associated to $W^{±}$, $Z^0$ production in $pp$ collisions.
The validity of eq. (\ref{eq7}) has been checked for several cases in $AA$ and $pp$ collisions \cite{salga,DiasdeDeus:1998du}.
Moreover, a different trend for the multiplicity distributions for interactions with and without charm, 
in agreement with our expectations, was already obtained more than twenty years ago by NA27 Collaboration \cite{Aguilar} in $pp$ collisions.
This difference was interpreted to be due to 
the more
central nature of collisions leading to charm
production.
This is also the case in our approach, 
when taking the shape given by eq. (\ref{eq7}).

From the experimental minimum bias distribution $P(n)$ measured by ALICE \cite{Abelev:2012rz} 
for the rapidity range $|\eta|<1.0$,
and using the relation established in eqs. (\ref{eq5}) and (\ref{eq6}) for the computation of
$\frac{n_{J/\psi}}{<n_{J/\psi}>}$ as a function of the multiplicity, 
one can calculate the probability distribution for the $J/\psi$, $P_c(n)$, accordingly to eq. (\ref{eq7}). 
Our results for the central rapidity region are shown in Fig. 2, compared to the 
minimum bias distribution $P(n)$ in this region. 
A particular shape, characterized by a reduction of the probability for low multiplicities and an increase 
at high multiplicities when compared to the minimum bias distribution is obtained.
Moreover, 
we have checked that,
when using PYTHIA 6.4 in the Perugia 2011 tune \cite{Skands:2010ak,Sjostrand:2006za} for the computation of
$\frac{n_{J/\psi}}{<n_{J/\psi}>}$ as a function of the multiplicity, accordingly to the result presented in
\cite{Abelev:2012rz,PorteboeufHoussais:2012gn} in the central rapidity region, and applying (\ref{eq7}), 
the opposite behaviour results\footnote{According to \cite{Abelev:2012rz}, 
the PYTHIA result exhibits a decrease of the $J/\psi$ multiplicity with
respect to the event multiplicity, contrary to data.}

\begin{figure*}[htb!]
\vskip -3cm
\begin{center}
\includegraphics[width=0.75\textwidth]{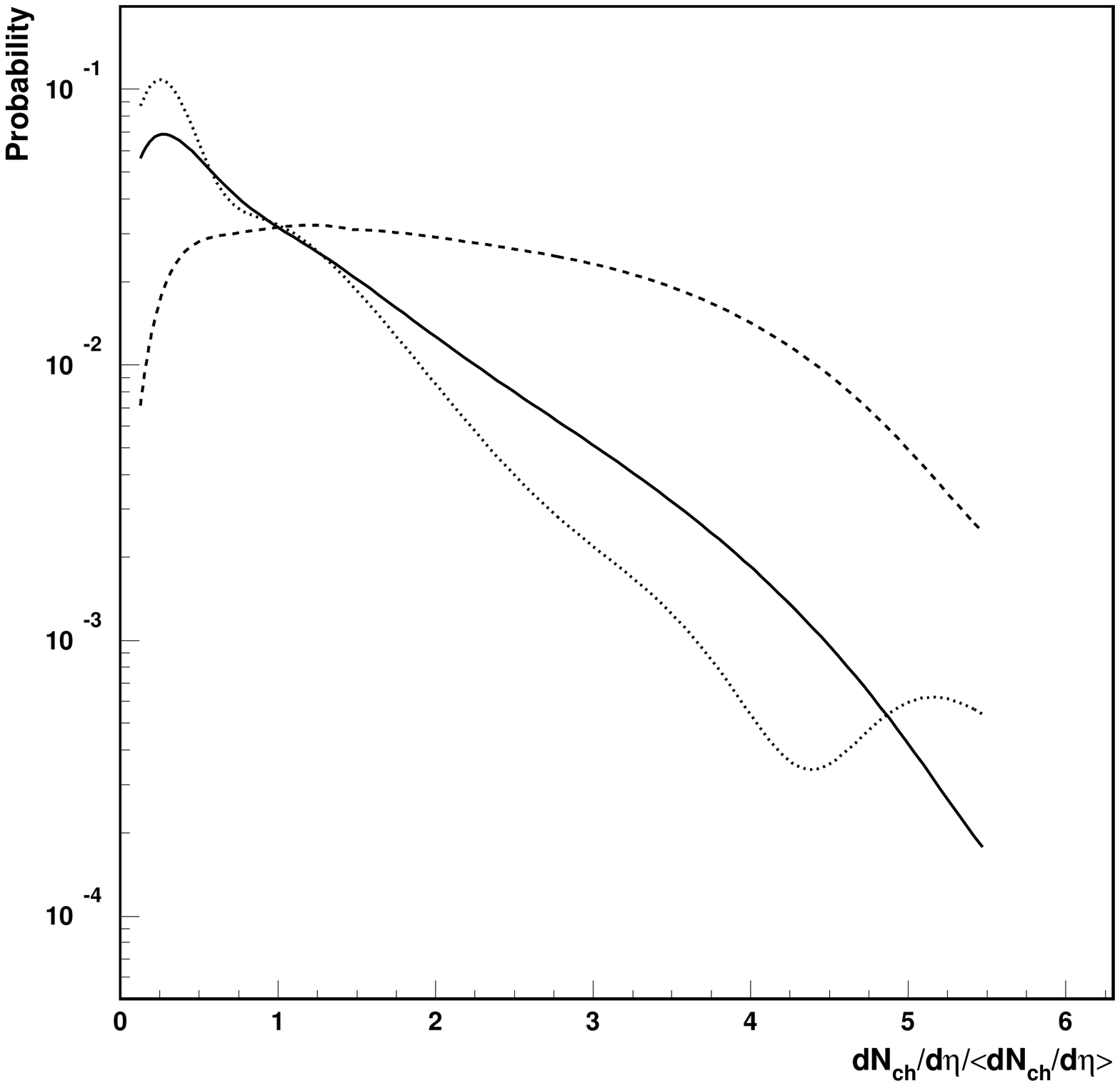}
\end{center}
\vskip -3cm
\caption{
Minimum bias experimental distribution $P(n)$ (solid line) compared to our obtained $P_c(n)$ distribution for the $J/\psi$ (dashed line) and to the $P_c(n)$ distribution obtained when using PYTHIA for the $J/\psi$ production (dotted line).}
\label{fig2}
\end{figure*}

\vskip 0.35cm
In conclusion, we have reproduced here the rise of 
$J/\psi$ production in the highest multiplicity events observed by the ALICE collaboration in $pp$ collisions. 
This increase, more pronounced in the central rapidity region, may be a consequence of the formation of a 
high density medium in $pp$ collisions at LHC energies. 
In this case, the linear dependence of $J/\psi$ production on the charged particle multiplicity obtained for low multiplicities 
--where the parton densities are smaller--, changes to an squared dependence 
when high multiplicity events are at play, 
due to the high string densities.

This behaviour can be checked by studying the probability distribution associated to the $J/\psi$ 
production, compared to the minimum bias probability distribution. A particular shape, characterized by a reduction of the probability for low multiplicities and an increase at high multiplicities when compared to the minimum bias distribution would be obtained. 

\vskip 0.35cm
{\it Acknowledgements.---}
We are grateful to N. Armesto, G. Martinez and C. A. Salgado
for useful
discussions.
This work is supported by
Ministerio de Economia y Competitividad 
of Spain (FPA2011-22776 and AIC-D-2011-0740) and FEDER.

\end{document}